\documentclass[sigconf]{acmart}
\usepackage{amsmath}     
\usepackage{algorithm}   
\usepackage{algpseudocode} 
\usepackage{bm}          
\usepackage{booktabs} 
\usepackage{multirow}
\usepackage{graphicx}
\usepackage{enumitem} 
\usepackage{subcaption}
\usepackage{makecell} 
\usepackage{balance}

\AtBeginDocument{%
  }

\copyrightyear{2026}
\acmYear{2026}
\setcopyright{cc}
\setcctype{by}
\acmConference[SIGIR '26]{Proceedings of the 49th International ACM SIGIR Conference on Research and Development in Information Retrieval}{July 20--24, 2026}{Melbourne, VIC, Australia}
\acmBooktitle{Proceedings of the 49th International ACM SIGIR Conference on Research and Development in Information Retrieval (SIGIR '26), July 20--24, 2026, Melbourne, VIC, Australia}
\acmISBN{979-8-4007-2599-9/2026/07}
\acmDOI{10.1145/3805712.3809630}
\algnewcommand\algorithmicinput{\textbf{Input:}}
\algnewcommand\Input{\item[\algorithmicinput]}
\algnewcommand\algorithmicoutput{\textbf{Output:}}
\algnewcommand\Output{\item[\algorithmicoutput]}
\begin{document}

\title{Beyond Dense Connectivity: Explicit Sparsity for Scalable Recommendation}

\author{Yantao Yu}
\email{yuyantao.yyt@alibaba-inc.com}
\affiliation{%
  \institution{Alibaba International Digital Commercial Group}
  \city{HangZhou}
  \country{China}
}

\author{Sen Qiao}
\email{qiaosen.qiaosen@alibaba-inc.com}
\affiliation{%
  \institution{Alibaba International Digital Commercial Group}
  \city{HangZhou}
  \country{China}
}

\author{Lei Shen}
\email{kenny.sl@alibaba-inc.com}
\affiliation{%
  \institution{Alibaba International Digital Commercial Group}
  \city{HangZhou}
  \country{China}
}

\author{Bing Wang}
\email{lingfeng.wb@alibaba-inc.com}
\affiliation{%
  \institution{Alibaba International Digital Commercial Group}
  \city{HangZhou}
  \country{China}
}

\author{Xiaoyi Zeng}
\email{yuanhan@alibaba-inc.com}
\affiliation{%
  \institution{Alibaba International Digital Commercial Group}
  \city{HangZhou}
  \country{China}
}

\renewcommand{\shortauthors}{Yantao Yu, Sen Qiao, Lei Shen, Bing Wang, and Xiaoyi Zeng}
\begin{abstract}
Recent progress in scaling large models has motivated recommender systems to increase model depth and capacity to better leverage massive behavioral data. However, recommendation inputs are high-dimensional and extremely sparse, and simply scaling dense backbones (e.g., deep MLPs) often yields diminishing returns or even performance degradation. 
Our analysis of industrial CTR models reveals a phenomenon of implicit connection sparsity: most learned connection weights tend towards zero, while only a small fraction remain prominent. 
This indicates a structural mismatch between dense connectivity and sparse recommendation data; by compelling the model to process vast low-utility connections instead of valid signals, the dense architecture itself becomes the primary bottleneck to effective pattern modeling.
We propose \textbf{SSR} (Explicit \textbf{S}parsity for \textbf{S}calable \textbf{R}ecommendation), a framework that incorporates sparsity explicitly into the architecture. SSR employs a multi-view "filter-then-fuse" mechanism, decomposing inputs into parallel views for dimension-level sparse filtering followed by dense fusion. Specifically, we realize the sparsity
via two strategies: a Static Random Filter that achieves efficient structural sparsity via fixed dimension subsets, and Iterative Competitive Sparse (ICS), a differentiable dynamic mechanism that employs bio-inspired competition to adaptively retain high-response dimensions. Experiments on three public datasets and a billion-scale industrial dataset from AliExpress (a global e-commerce platform) show that SSR outperforms state-of-the-art baselines under similar budgets. Crucially, SSR exhibits superior scalability, delivering continuous performance gains where dense models saturate. The code is available at \url{https://github.com/Atticus666/SSRNet}.
\end{abstract}




\maketitle

\section{Introduction}
Deep learning recommender systems (DLRS) are the core ranking engines in many online services. Inspired by the success of LLMs \cite{kaplan2020scaling,hoffmann2022training}, we investigate whether recommender models exhibit similar scaling properties, where performance improves as model capacity and data size grow together. In practice, mainstream industrial CTR backbones such as Wide\&Deep \cite{cheng2016wide} and DLRM \cite{naumov2019deep} remain relatively shallow, often 3–4 layers. Attempts to simply scale up these dense MLP-based architectures frequently lead to diminishing returns or even performance degradation, as reported in prior studies \cite{rendle2020neural,liu2020autofis}. This implies that naive scaling of dense architectures is suboptimal.
\begin{figure}[!t]
  \centering
  \includegraphics[width=\linewidth]{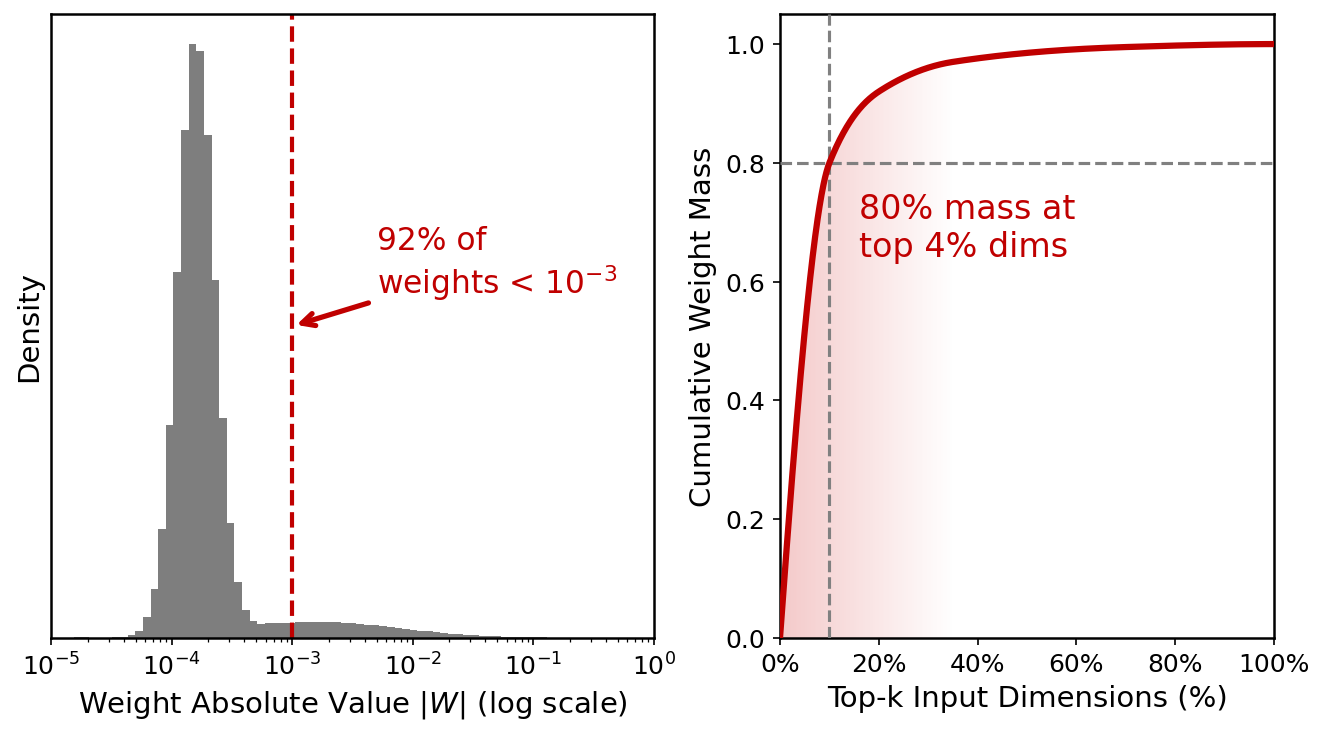} 
  \caption{Sparsity analysis of the hidden layer in online CTR backbone.  (Left) 92\% of weights are suppressed to near-zero $(< 10^{-3})$. (Right) 80\% of weight power concentrates in the top 4\% of dimensions.}
  \Description{Sparsity analysis in dense CTR backbones.}
  \label{fig:1}
\end{figure}

A fundamental issue is the mismatch between dense connectivity and sparse recommendation data. Unlike images or text with natural spatial or sequential locality, recommendation inputs consist of hundreds of heterogeneous feature fields where only a small subset is relevant for any given sample \cite{zhang2021deep,pi2020search}. As we show in Figure \ref{fig:1}, learned weights in a production CTR model exhibit extreme implicit sparsity: 92\% of connections are suppressed to near-zero, and 80\% of weight mass concentrates in just 4\% of dimensions. This confirms that the dense architecture itself becomes the bottleneck to effective scaling, and we provide a detailed analysis and theoretical grounding in Section 2.

Building on this insight, we propose \textbf{SSR} (Explicit \textbf{S}parsity for \textbf{S}calable \textbf{R}ecommendation),  a framework tailored for scaling on sparse recommendation data. SSR introduces a paradigm shift from implicit
weight suppression to explicit signal filtering, building on a simple
principle: first filter, then fuse. It performs explicit dimension-level sparse filtering before dense nonlinear fusion. We realize the sparsity via two complementary strategies: a Static Random Filter (SSR-S) that achieves efficient structural sparsity via fixed dimension subsets at zero FLOP cost, and Iterative Competitive Sparse (ICS/SSR-D), a differentiable dynamic mechanism that introduces sparsity to adaptively filter dimensions based on sample context. Unlike existing methods that rely on soft attention \cite{huang2019fibinet} or post-hoc pruning \cite{liu2020autofis}, which maintain a fully connected graph and thus fail to block noise at scale, SSR enforces explicit sparsity,
preventing noise propagation at the source and thus providing a cleaner gradient flow for scaling. The main contributions of this paper are summarized as follows:
\begin{itemize}
\item We analyze the problem of scaling dense MLPs on sparse data, highlighting that implicit weight suppression fails to block noise, and provide evidence of strong sparse connection in Figure \ref{fig:1}.
\item We propose SSR, shifting the paradigm from implicit weight suppression to explicit signal filtering. It realizes explicit sparsity to isolate noise before dense interaction, ensuring expanded capacity is dedicated to valid signals.
\item We introduce two strategies to realize the explicit sparsity: a Static Random Filter for efficient structural sparsity, and ICS, a differentiable dynamic filtering mechanism that enables input-adaptive sparsification to capture complex dependencies.

\item Experiments on three public datasets and a billion-scale industrial dataset from AliExpress demonstrate that SSR achieves better accuracy under comparable compute budgets and exhibits more stable improvements when scaling size.
\end{itemize}

\section{Motivation and Theoretical Foundation}
Before presenting the SSR framework, we provide the analytical and theoretical motivation underlying our design choices. 

\subsection{Why Dense MLPs Fail for Recommendation}
Recommendation inputs differ fundamentally from language or vision data. They are high-dimensional yet extremely sparse \cite{zhang2021deep,wang2025towards,kasalicky2025future}, each instance typically activates only a small subset of informative dimensions in a large feature space \cite{pi2020search}. Unlike images or text, where inputs exhibit natural spatial or sequential locality that architectures like CNNs and Transformers. Recommendation inputs consist of hundreds of heterogeneous feature fields such as user profiles, item attributes, contextual signals, and behavioral sequences, concatenated into a flat vector with no inherent adjacency or ordering among dimensions. For a specific impression or purchase, only a few contextual signals and historical preferences are truly relevant, while the vast majority are weakly relevant for that specific sample \cite{yu2019input,lu2021dual}. This sparsity pattern means the effective response of the model (e.g., weight mass) is concentrated on a small fraction of the input dimensions.

In contrast, a fully connected layer enforces globally dense connectivity by coupling each output neuron with all input dimensions. This forces the model to process vast low-utility connections, which dilutes valid signals and burdens the optimizer with suppressing noise rather than learning complex patterns \cite{song2019autoint}. We argue that this constitutes a \textit{misalignment of inductive bias}: the dense connectivity assumes all dimension pairs are equally likely to interact, whereas the data exhibits highly concentrated, subset-based interactions.

To support this analysis, we visualize the learned weights of the fully connected layer in an online industrial CTR model (Figure \ref{fig:1}). This model was trained without any sparsity-inducing constraints (e.g., L2 regularization). Despite its dense design, the learned weights exhibit a highly sparse operating pattern: more than 92\% of connections are implicitly suppressed to near-zero values ($< 10^{-3}$), and 80\% of the weight mass is concentrated in the top 4\% of the input dimensions. While this confirms a strong sparsity preference driven by the data distribution, such implicit suppression is inefficient: many weights are simply driven close to zero, which neither eliminates the interference of noise nor provides a mechanism for signal filtering. Making this sparsity explicit \cite{fedus2022switch,frankle2018lottery}, i.e., transforming it from an implicit training artifact into a controllable architectural design, is therefore the key to overcoming the scaling bottleneck. However, what constitutes noise varies across users, so a static sparse structure shared by all samples misses the context dependence of recommendation. To scale effectiveness, we need both structural and dynamic, sample-conditional sparsity.

\subsection{Sparsity as Inductive Bias Alignment}
From an inductive bias perspective, a model architecture implicitly encodes assumptions about the structure of its input data. CNNs encode spatial locality through convolutional kernels; Transformers encode sequential dependencies via self-attention \cite{tolstikhin2021mlp}. Both succeed because their structural design matches the natural structure of the data.

Recommendation inputs, however, lack such natural locality. Hundreds of heterogeneous feature fields are concatenated into a flat vector with no inherent spatial or temporal ordering. Dense MLP's fully connected topology imposes no structural prior, treating all dimension pairs as equally likely to interact. When the data's effective interactions are concentrated on small feature subsets, as our empirical analysis confirms (Figure \ref{fig:1}). This design becomes a misaligned inductive bias, forcing the optimizer to expend most of its capacity learning which connections to suppress rather than which patterns to model.

This perspective is supported by recent theoretical work on structured sparsity. Hayase and Karakida \cite{hayase2024understanding} proved that MLP-Mixer \cite{tolstikhin2021mlp} is mathematically equivalent to a \textit{wide and sparse} MLP. Its Token-Mixing and Channel-Mixing layers can be expressed via Kronecker product structure:
\begin{equation}
\text{vec}(\mathbf{W}\mathbf{X}) = (\mathbf{I}_C \otimes \mathbf{W})\, \text{vec}(\mathbf{X})
\label{eq:token-mixing}
\end{equation}
\begin{equation}
\text{vec}(\mathbf{X}\mathbf{V}) = (\mathbf{V}^\top \otimes \mathbf{I}_S)\, \text{vec}(\mathbf{X})
\label{eq:channel-mixing}
\end{equation}
This structure yields an effective width of $m = S \times C$ (potentially $10^4$--$10^6$) while maintaining a non-zero weight ratio of only $1/C$ or $1/S$, which is inherently highly sparse. Moreover, this Kronecker product parameterization carries an implicit $L_1$ regularization effect. Together with the Golubeva hypothesis \cite{golubevawider}, which states that at fixed parameter count, increasing width (and thus sparsity) consistently improves generalization, these findings provide theoretical grounding for why structured sparsity serves as a beneficial inductive bias.

SSR extends this principle from vision to recommendation. While MLP-Mixer relies on fixed mathematical structure (Kronecker products) that exploits the spatial regularity of image patches, recommendation data lacks such regularity: which feature interactions are informative is highly data-dependent and sample-dependent. This motivates SSR's design of two complementary explicit sparsity mechanisms, namely static random filtering for efficient structural sparsity and dynamic competitive filtering for sample-adaptive selection. The transition from MLP-Mixer to SSR thus represents a advance: from implicit sparsity embedded in fixed mathematical structure to explicit sparsity engineered for the data's intrinsic properties.
this motivates the filter-then-fuse paradigm detailed in the next section.


\section{The SSR Framework}
\begin{figure*}[t]
  \centering
  \includegraphics[width=\textwidth]{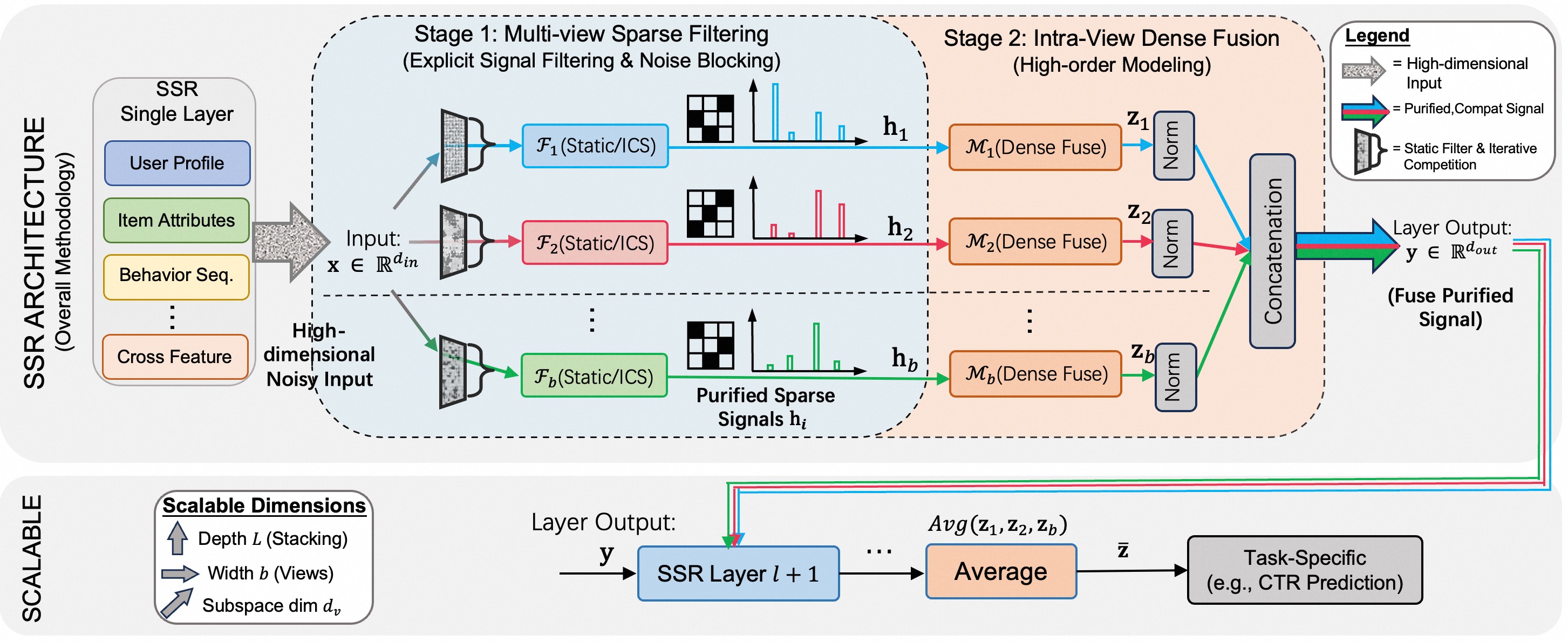}
  \caption{The SSR Framework: Explicit Sparsity for Scalable Recommendation.}
  \Description{The SSR Framework: Explicit Sparsity for Scalable Recommendation.}
  \label{fig:2}
\end{figure*}

We propose SSR (Explicit Sparsity for Scalable Recommendation) framework to resolve the mismatch between globally dense connectivity and sparse input data. 
In this section, we detail the design of a single SSR Layer, which comprises two cascaded stages: (1) Multi-view Sparse Filtering and (2) Intra-view Dense Fusion. Figure \ref{fig:2} presents an overview of the framework.

\subsection{Overview}
To overcome the scaling issue caused by the indiscriminate mixing and signal dilution inherent in traditional densely connected layers, SSR introduces a new computational paradigm based on explicit signal filtering.
First, the model converts raw features—including user profiles, candidate item attributes, cross-feature statistics, and behavior sequences—into embeddings. These embeddings are concatenated to form the initial input vector $\mathbf{x} \in \mathbb{R}^{d_{\text{in}}}$.

Unlike a standard dense layer that learns a global mapping $\mathbf{x} \in \mathbb{R}^{d_{in}}$, SSR decouples the modeling task into $b$ independent purification views. For each view $i \in \{1,\dots,b\}$, we define a view-specific mapping $\phi_i$ that processes the input into a local subspace representation $\mathbf{z}_i \in \mathbb{R}^{d_v}$. 
Each mapping filters the key dimensions in the full input $\mathbf{x}$ and projects them into a low-dimensional subspace.
Each mapping $\phi_i$ is implemented through a strict two-stage process: Sparse Filtering ($\mathcal{F}_i$) to filter the information, followed by Dense Fusion ($\mathcal{M}_i$) to process it. The view outputs $\{\mathbf{z}_1, \dots, \mathbf{z}_b\}$ are then aggregated to form the layer output; the specific aggregation strategy differs between intermediate and final layers.

\subsection{Multi-view Sparse Filtering}

This stage constitutes the "Filter" stage of the SSR framework, implementing strict dimension-level signal filtering. We define a set of sparse filter operators $\{\mathcal{F}_1, \dots, \mathcal{F}_b\}$. For the $i$-th view, the operator extracts a purified representation $\mathbf{h}_i \in \mathbb{R}^{d_v}$ from the high-dimensional input $\mathbf{x}$:
\begin{equation}
\mathbf{h}_i = \mathcal{F}_i(\mathbf{x}) \label{eq:2}
\end{equation}
This process essentially performs $b$ parallel filtering operations. We propose two instantiation strategies for $\mathcal{F}_i$, trading off efficient structural sparsity and context-aware dynamic sparsity.

\noindent \textbf{SSR-S: Static Random Filter (Static Instantiation)}
This strategy treats $\mathcal{F}_i$ as a sample-agnostic operator to enforce structural sparsity. We implement $\mathcal{F}_i$ using a binary selection matrix $M_i \in \{0,1\}^{d_{in} \times d_v}$, where each column is strictly a one-hot vector. Furthermore, this matrix remains fixed after initialization.
To construct $M_i$, we sample $d_v$ feature indices from the input dimension range $\{1, \dots, d_{in}\}$. The sampling is performed uniformly without replacement within each view, ensuring distinct features in a single subspace. However, the sampling is independent across different views, allowing feature overlap. This independence creates a "Feature Bagging" effect \cite{breiman2001random}, promoting structural diversity and robustness across the parallel views. The filtered feature is calculated as:
\begin{equation}
\mathbf{h}_i = \mathbf{x} \mathbf{M}_i \label{eq:3}
\end{equation}
Since $M_i$ consists of column-wise one-hot vectors, the operation is implemented not as a matrix multiplication, but as a zero-FLOP parallel gather operation (i.e., direct index slicing). This blocks the propagation of unselected dimensions at the source.

Existing methods like Statistical Top-k \cite{you2025spark} or even our own dynamic ICS utilize logical sparsity: they multiply non-informative features by zero, but the physical computation graph remains wide ($O(d^2)$). In contrast, SSR-S enforces hard dimension reduction. By strictly slicing the input indices before computation, it decouples the dimension selection cost from the inference cost.

\noindent\textbf{SSR-D: Iterative Competitive Sparse (Dynamic Instantiation)}
To capture context-aware dependencies, we employ ICS (detailed in Sec. \ref{sec3}), a dynamic mechanism. ICS dynamically adjusts the focus based on the input's semantic context. It sparsifies the input by  actively zeroing out less salient elements in the input vector $\mathbf{x}$ while retaining high-response values. The formula for $\mathbf{h}_i$ becomes:
\begin{equation}
\mathbf{h}_i = \text{ICS}_i(\mathbf{x}\mathbf{W}^{proj}_{\text{i}}) \label{eq:4}
\end{equation}
Here, $\mathbf{h}_i \in \mathbb{R}^{d_{v}^*}$, where the view dimension is typically expanded (e.g., $d_{v}^* > d_{v}$) to maintain capacity for adaptive dimension sparsity, unlike the static strategy. $\mathbf{W}^{proj}_{i} \in \mathbb{R}^{d_{\text{in}} \times d_{v}^*}$ is a learnable projection matrix for view $i$. The output $\mathbf{h}$ is a sparse representation, in the $d_{v}^*$-dimensional space, where most non-critical elements are strictly truncated to zero. 

\subsection{Intra-view Dense Fusion}
Following dimension-level sparse filtering, the input has been distilled into $b$ views of purified vectors $[h_1, ..., h_b]$. While the first stage blocks noise, this second stage focuses on exploiting this sparsity to enable efficient high-order modeling within a refined signal environment. Its strategic application exclusively within the refined subspaces prevents the re-aggregation of low-utility connections, resolving the signal dilution issue inherent in globally dense architectures.

Mathematically, this operation is equivalent to applying a Block-Diagonal weight matrix $\mathbf{W}_{\text{block}} = \text{diag}(\mathbf{V}_1, \dots, \mathbf{V}_b)$ to the concatenated input. Unlike a standard dense layer where all dimensions interact, the block-diagonal structure enforces strict semantic isolation between views, ensuring that features from the $i$-th view are transformed exclusively by parameters $\mathbf{V}_i \in \mathbb{R}^{d_{v} \times d_{v}}$ for static or $V_i \in \mathbb{R}^{d_{v}^* \times d_{v}}$ for dynamic. In practice, this is efficiently implemented as b parallel projections, avoiding the storage of zero-valued off-diagonal blocks.
The output $\mathbf{z}_i$ for the $i$-th view is calculated as:
\begin{equation}
\mathbf{z}_{i} = \sigma\left( \mathbf{h}_i \mathbf{V}_i + \mathbf{bias}_i \right) \label{eq:5}
\end{equation}
Here, $\sigma$ is an activation function (such as GELU). For intermediate layers, the outputs from all views are processed with Layer Normalization and recombined via concatenation:
\begin{equation}
\mathbf{y} = \text{concat}(\text{LayerNorm}(\mathbf{z}_1),...,\text{LayerNorm}(\mathbf{z}_b)) \in \mathbb{R}^{b \cdot d_{v}} \label{eq:6}
\end{equation}
The number of parameters in this structure is $O(b \cdot d_{v}^2)$.
Compare this with a standard fully connected layer, which has a complexity of $O((b \cdot d_{v})^2)$. By utilizing the independence of each view, SSR reduces complexity by a factor of $1/b$. This makes it possible to significantly expand parameters within the same computational budget.


\subsection{Last-Layer Aggregation}\label{sec:scalable}

In intermediate layers, view outputs are concatenated and passed to the next layer as defined in Eq.~\eqref{eq:6}. However, for the final layer that produces prediction logits (e.g., CTR/CVR scores), the aggregation strategy switches from concatenation to averaging:
\begin{equation}
\bar{\mathbf{z}} = \frac{1}{b}\sum_{i=1}^{b} \text{LayerNorm}(\mathbf{z}_i) \label{eq:last_layer}
\end{equation}
Each view independently completes dense fusion to obtain $\mathbf{z}_i$, and the resulting shared representation $\bar{\mathbf{z}}$ is then fed into task-specific prediction heads via fully connected layers, e.g., 
\begin{equation}
y_{\text{ctr}} = \sigma(\mathbf{W}_{\text{ctr}}\bar{\mathbf{z}} + \mathbf{b}_{\text{ctr}}) \label{eq:pred}
\end{equation}
Averaging offers two advantages over concatenation. First, it pushes all views toward a shared semantic space instead of letting them drift independently. Concatenation preserves discrepancies between views, but averaging encourages consistency. Second, averaging fixes the prediction head input dimension at $d_v$ regardless of the view count. With concatenation, the dimension would grow to $b \cdot d_v$, making the prediction head scale linearly with the number of views.

\section{Iterative Competitive Sparse}\label{sec3}
As the core mechanism for dynamic instantiation in SSR, Iterative Competitive Sparse (ICS) is a differentiable operator that differs from traditional sparsification—typically handled by discrete Top-K sorting—as a continuous dynamical system. This formulation
enables end-to-end, adaptive sparse filtering across dimensions.

We treat the input $\mathbf{p} \in \mathbb{R}^{d_{v}}$ as a population in an ecosystem where feature intensities represent vitality.
This framework redefines sparsification as a discrete-time nonlinear dynamical system rather than a static sorting task. It comprises three continuous stages: initialization, iterative competition, and signal recovery. The ICS forward pass is fully differentiable, enabling integration into gradient-based optimization. 
The standard flow is shown in Algorithm \ref{alg:ics}.
\begin{algorithm}[t]
    \caption{Iterative Competitive Sparse (ICS) Forward Pass}
    \label{alg:ics}
    \begin{algorithmic}[1] 
        \Input Project feature $\mathbf{z}  \in \mathbb{R}^{d_{v}}$, Iterations $T$,  Learnable extinction rates $\{\alpha_t\}_{t=0}^{T-1}$, Learnable scale $\gamma$
        \Output Sparse feature vector $\mathbf{y} \in \mathbb{R}^{d_{v}}$
        
        \State \textbf{Initialize:} $\mathbf{x}^{(0)} \leftarrow \operatorname{ReLU}(\mathbf{z})$
        
        \For{$t = 0 \text{ to } T - 1$}
            \State $\mu^{(t)} \leftarrow \operatorname{Mean}(\mathbf{x}^{(t)})$
            \State $\mathbf{x}^{(t+1)} \leftarrow \operatorname{ReLU}(\mathbf{x}^{(t)} - \alpha_t \cdot \mu^{(t)})$
        \EndFor

        
        \State \textbf{Signal Recovery:} $\mathbf{y} \leftarrow \gamma \odot \mathbf{x}^{(T)}$
        
        \State \textbf{return} $\mathbf{y}$
    \end{algorithmic}
\end{algorithm}
\subsection{Initialization and Competitive Dynamics}

Dynamic competition requires feature intensity to have a non-negative physical meaning. Therefore, we first rectify the input to be non-negative. We define the initial system state as:
\begin{equation}
\mathbf{x}^{0} = \text{ReLU} (\mathbf{z})\label{eq:7}
\end{equation}

Then, the system enters an iterative process for $T$ rounds ($t = 0, \dots, T-1$). 
During iterations, a mean-field global inhibition force drives features toward extinction.
We define the global inhibition field $\mu^{(t)}$ in step $\lambda_t$ as the mean of all current features.

\begin{equation}
\mu^{(t)} = \frac{1}{d_{v}} \sum_{j=1}^{d_{v}} x_j^{(t)}\label{eq:9}
\end{equation}

The state update follows the "survival of the fittest" rule. Only features significantly stronger than the inhibition field can survive. The rest will converge to true zero as hard sparsity. The specific update equation is:
\begin{equation}
\mathbf{x}^{(t+1)} = \text{ReLU}\left( \mathbf{x}^{(t)} - \alpha_t \cdot \mu^{(t)} \right)\label{eq:10}
\end{equation}
Here, $\mathbf{\alpha} = \{\alpha_0, \dots, \alpha_{T-1}\} ,\alpha_t \in \mathbb{R}$, we introduce $T$ learnable extinction rates where different iterations use different $\alpha_t$. Crucially, the iterative design ($T>1$) is necessary because the statistical distribution of the features is not stable during the iterative process. A single-step thresholding ($T=1$) relies on a static estimation of the noise floor. Through $T$ iterations, as noise is progressively extinguished, the mean $\mu^{(t)}$ is continuously refined to reflect the true signal baseline. This allows the model to perform progressive filtering by removing coarse noise first and fine-tuning later, thereby approximating a complex non-linear sparsification that a single linear filtering cannot achieve.

In each iteration, we only perform additions/subtractions and compute the mean, all of which are $O(N)$ operations. Over $T$ iterations, the total complexity is $O(T\cdot N)$. Since $\alpha_t > 0$ and $\mu^{(t)} \ge 0$, the update rule ensures that no feature intensity can increase. The system forms a monotonically non-increasing sequence:
\begin{equation}
|\mathbf{x}^{(t+1)}|_1 \le |\mathbf{x}^{(t)}|_1 \label{eq:11}
\end{equation}
This inequality implies that the total energy of the system inevitably decays over time $t$. While this effectively filters out noise, it also causes significant attenuation of the useful signal intensity. 

\subsection{Signal Recovery }
To counteract this inherent attenuation, we introduce a learnable scale parameter $\gamma$.
After $T$ rounds of iteration, the sparse state $\mathbf{x}^{(T)}$ is mapped to the final output $\mathbf{y}$ through a linear transformation:

\begin{equation}
\mathbf{y} = \gamma \odot \mathbf{x}^{(T)}\label{eq:12}
\end{equation}
we introduce $\gamma$ as a learnable rescaling parameter, we implement $\gamma \in \mathbb{R}^{d_{v}}$ as a vector, assigning an independent weight to each dimension.
While theoretically, the subsequent linear layer could absorb a scalar multiplication, we specifically introduce  $\gamma$ to decouple recovery from  transformation. The parameter 
$\gamma$ serves as a variance stabilizer, ensuring numerical stability and an optimal dynamic range for the optimization process.

\subsection{Comparison with Other Top-k Mechanisms}
Our ICS mechanism offers distinct advantages over existing differentiable selection strategies. First, compared to Straight-Through Estimator (STE) \cite{bengio2013estimating} based Top-k methods, ICS eliminates the gradient mismatch problem. By formulating sparsification as a continuous dynamical system rather than a discrete truncation, ICS ensures a consistent gradient flow that stabilizes training.
Second, unlike Soft Top-k relaxations or NeuralSort \cite{grover2019stochastic}, which typically involve sorting operations with super-linear $O(N\log N)$ complexity, ICS achieves sparsity through parallel competitive inhibition. This results in a strictly linear $O(T\cdot N)$ complexity, avoiding the computational bottleneck of sorting high-dimensional recommendation features while ensuring that noise dimensions are driven to true zero rather than merely assigned low probabilities.
\section{Experiments}
This section aims to address the following core research questions:

\begin{enumerate}[label=\textbf{RQ\arabic*}]
\item \textbf{Effectiveness \& Efficiency:}  Does SSR outperform SOTA models on mainstream benchmarks in terms of both prediction accuracy and computational efficiency?

\item \textbf{Scalability:} Does SSR scale effectively, i.e., does performance consistently improve as the model scale increases?

\item \textbf{Ablation \& Mechanism:} What are the respective contributions of the sparse filtering design and the dense fuse? Does ICS truly achieve dynamic sparsity?

\item \textbf{Online A/B Tests:} Does deploying SSR online yield  significant lifts in key business metrics  under  latency constraints?
\end{enumerate}

\subsection{Experimental Setup }
\subsubsection{Datasets}
We evaluated on a large-scale industrial dataset and three public datasets, e.g., Criteo\footnote{\url{https://www.kaggle.com/c/criteo-display-ad-challenge/data}}, Avazu\footnote{\url{https://www.kaggle.com/c/avazu-ctr-prediction}}, Alibaba\footnote{\url{https://tianchi.aliyun.com/dataset/408}}. Dataset statistics are summarized in Table 1.
The industrial dataset contains over 1 billion production logs from AliExpress, a global cross-border e-commerce platform under Alibaba International Digital Commerce Group. The data is collected from its recommendation system, which serves personalized product recommendations to users worldwide. The dataset encompasses more than 300 feature fields including user profiles, item attributes, and contextual signals, and we use a time-based split where the most recent day is used for validation and testing to mimic an online setting.
For public datasets, we follow the standard random split (8:1:1).
All numerical features are log-transformed and discretized, and categorical features with frequency $\le 5$ are removed.

\begin{table}[t]
    \centering
    \caption{Statistics of Datasets}
    \label{tab:dataset_stats}
    \resizebox{\linewidth}{!}{%
    \begin{tabular}{lrrrrr}
        \toprule
        \textbf{Data} & \textbf{\#Samples} & \textbf{\makecell[r]{\#Positive\\ Ratio}} & \textbf{\makecell[r]{\#Categorical\\ Features}} & \textbf{\makecell[r]{\#Numerical\\ Features}} & \textbf{\makecell[r]{\#Feature\\ Values}} \\
        \midrule
        Avazu      & 40,428,967      & 16.98\% & 23  & 0   & 1,544,489 \\
        Criteo     & 45,840,617      & 25.62\% & 26  & 13  & 998,974   \\
        Alibaba    & 42,299,905      & 3.89\%  & 23  & 4  & 1,342,817 \\
        Industrial & 1,003,204,206   & \makecell[r]{3.45\% \\ / 0.08\% } & 183 & 129  & --        \\
        \bottomrule
    \end{tabular}%
}
\end{table}

\subsubsection{Evaluation protocols.}
To evaluate the proposed method, we consider both prediction effectiveness and computational efficiency. In terms of effectiveness, we use AUC and LogLoss across all datasets. For the industrial datasets, we additionally introduce GAUC to mitigate user activity bias and focus on intra-user ranking performance. The industrial dataset includes two tasks, \emph{click} and \emph{pay}; for the \emph{pay} task, we evaluate on the full sample space. Regarding efficiency and scalability, we report Params and FLOPs. Note that the parameter count includes only the backbone network (excluding embedding tables) to decouple architectural evaluation from dataset-specific feature cardinality. Furthermore, we calculate FLOPs based on a single inference pass of the neural network components to serve as a proxy for the computational overhead during the training phase.

\subsubsection{Baselines.}
We benchmark SSR against four groups of representative methods: (1) Classic Deep Models: DeepFM \cite{guo2017deepfm} and DCN v2 \cite{wang2021dcn}, which serve as standard baselines utilizing dense feature interactions; (2) Attention-based \& Dynamic models: AutoInt \cite{song2019autoint} and MMOE \cite{ma2018modeling}, which employ self-attention or gating mechanisms for adaptive feature learning; (3) Feature Selection (AutoML): AutoFIS \cite{liu2020autofis}, AFN \cite{cheng2020adaptive}, a state-of-the-art method that improves efficiency by pruning redundant interactions; and (4) SOTA scalable architectures: Wukong \cite{zhang2024wukong} and RankMixer \cite{zhu2025rankmixer}, representing the latest advancement in high-performance industrial recommendation.

All methods are implemented in TensorFlow and trained on a NVIDIA A100 cluster. For a fair comparison, we set the embedding dimension to 16 for all models and used Adam with a batch size of 1024 and early stopping. The iterations $T$ of ICS are set to 5, the learnable extinction rates $\alpha_t$ are initialized to 0.1, and the Learnable scale $\gamma$ is initialized to a vector of 1.

\subsection{Effectiveness \& Efficiency (RQ1)}\label{rq:1}
\subsubsection{Performance on Industrial Datasets}
Table \ref{tab:inds} presents quantitative results for Click and Pay tasks on industrial datasets. We compare three groups of baselines against the static random strategy SSR-S and the dynamic ICS strategy SSR-D. SSR consistently outperforms classic feature interaction models. For instance, the static SSR-S variant achieves a Click AUC of 0.6644, surpassing standard baselines like DeepFM and DCN v2. Notably, SSR-S outperforms a Dense MLP of comparable parameter size, indicating that the performance gain comes from the sparse architecture itself rather than parameter capacity.

\begin{table}[t]
\centering
\caption{Overall performance and efficiency comparison on the industrial dataset. The best results are highlighted in bold. * indicates statistical significance over the best baseline with $p < 0.05$.}
\label{tab:inds}
\resizebox{\linewidth}{!}{%
\begin{tabular}{lcccccc}
\toprule
\multirow{2}{*}{\textbf{Model}} & \multicolumn{2}{c}{\textbf{CLICK}} & \multicolumn{2}{c}{\textbf{PAY}} & \multirow{2}{*}{\textbf{\#Params}} & \multirow{2}{*}{\textbf{FLOPs/1}} \\ 
\cmidrule(lr){2-3} \cmidrule(lr){4-5} 
 & \textbf{AUC} & \textbf{GAUC} & \textbf{AUC} & \textbf{GAUC} & & \\ 
\midrule
Dense MLP & 0.6593 & 0.6281 & 0.8083 & 0.6770 & 60M & 3.4G \\
DeepFM & 0.6563 & 0.6251 & 0.8053 & 0.6730 & 13M & 0.6G \\
DCN v2 & 0.6571 & 0.6262 & 0.8065 & 0.6742 & 15M & 0.9G \\
MMoE & 0.6578 & 0.6267 & 0.8063 & 0.6757 & 21M & 1.2G \\
AutoInt & 0.6594 & 0.6279 & 0.8078 & 0.6769 & 26.2M & 1.7G \\
AutoFIS* & 0.6592 & 0.6285 & 0.8085 & 0.6777 & \textbf{10.8M} & \textbf{0.5G} \\
Wukong & 0.6615 & 0.6298 & 0.8115 & 0.6805 & 93M & 2.9G \\
RankMixer & 0.6621 & 0.6305 & 0.8122 & 0.6815 & 101M & 3.2G \\
\midrule
\textbf{SSR-S} & 0.6644 & 0.6326 & 0.8162 & 0.6841 & 57M & 1.4G \\
\textbf{SSR-D} & \textbf{0.6667}$^{*}$ & \textbf{0.6351}$^{*}$ & \textbf{0.8194}$^{*}$ & \textbf{0.6862}$^{*}$ & 100M & 3.3G \\
\bottomrule
\multicolumn{7}{l}{\footnotesize * For AutoFIS, metrics refer to the re-training phase (post-pruning).} \\
\end{tabular}%
}
\end{table}


In comparisons with automated and attention-based models, although AutoFIS benefits from a low parameter count during retraining, its limited capacity results in a suboptimal AUC of 0.6592. Similarly, AutoInt incurs higher computational costs of 1.7G FLOPs compared to 1.4G for SSR-S yet yields a lower score of 0.6594. These self-attention mechanisms use softmax to assign strictly positive weights ($\alpha_{ij} > 0$) to all feature pairs, thereby preserving a fully connected graph similar to that of a dense fully connected layer.

Against state-of-the-art architectures, the dynamic SSR-D variant achieves the highest overall performance. While RankMixer serves as the strongest baseline with a Click AUC of 0.6621, SSR-D surpasses it across all metrics, reaching 0.6667 in Click AUC and 0.8194 in Pay AUC. Finally, SSR offers a superior efficiency trade-off. SSR-S outperforms RankMixer using only 56\% of the parameters and 44\% of the FLOPs, validating the benefits of structured sparsity. SSR-D operates within a similar computation budget to RankMixer but delivers significant performance gains, verifying the effectiveness of the Iterative Competitive Sparse mechanism.

\subsubsection{Generalization on Public Benchmarks}

\begin{table*}[htbp]
\centering
\caption{Performance and efficiency comparison on public benchmarks (Avazu, Alibaba, and Criteo). The best results are highlighted in bold. * indicates statistical significance over the best baseline with $p < 0.05$.}
\label{tab:public_benchmarks}
\resizebox{\textwidth}{!}{%
\begin{tabular}{l|cccc|cccc|cccc}
\toprule
\multirow{2}{*}{\textbf{Model}} & \multicolumn{4}{c|}{\textbf{Avazu}} & \multicolumn{4}{c|}{\textbf{Alibaba}} & \multicolumn{4}{c}{\textbf{Criteo}} \\
\cmidrule(lr){2-5} \cmidrule(lr){6-9} \cmidrule(lr){10-13}
 & \textbf{AUC} & \textbf{LogLoss} & \textbf{Params*} & \textbf{FLOPs} & \textbf{AUC} & \textbf{LogLoss} & \textbf{Params*} & \textbf{FLOPs} & \textbf{AUC} & \textbf{LogLoss} & \textbf{Params*} & \textbf{FLOPs} \\
\midrule
DeepFm & 0.7752 & 0.3801 & 0.23M & 464.2 M & 0.6594 & 0.1604 & 0.21M & 421.3 M & 0.7986 & 0.4539 & 0.29M & 599.5 M \\
DCN v2 & 0.7729 & 0.3915 & 0.36M & 736.8 M & 0.6526 & 0.1659 & 0.35M & 712.0 M & 0.8064 & 0.4537 & 0.69M & 1.22 G \\
AFN & 0.7755 & 0.3839 & 0.15M & 317.4 M & 0.6757 & 0.1638 & 0.15M & 316.4 M & 0.8080 & 0.4561 & 0.90M & 1.96 G \\
AutoInt & 0.7722 & 0.3859 & 0.07M & 850.2 M & 0.6784 & 0.1575 & 0.29M & 1.18 G & 0.8053 & 0.4462 & 0.01M & 1.66 G \\
AutoFIS* & 0.7802 & 0.3792 & 0.23M & 472.7 M & 0.6637 & 0.1602 & 0.21M & 427.2M & 0.8089 & 0.4430 & 0.23M & 472.7 M \\
Wukong & 0.7756 & 0.3826 & 0.17M & 719.6 M & 0.6782 & 0.1567 & 0.17M & 694.6 M & 0.8073 & 0.4445 & 0.18M & 799.7 M \\
RankMixer & 0.7772 & 0.3818 & 0.64M & 1.32 G & 0.6801 & 0.1566 & 0.63M & 1.30 G & 0.8092 & 0.4427 & 1.15M & 2.36 G \\
\midrule
\textbf{SSR-S} & 0.7827* & 0.3781 & \textbf{0.33M} & \textbf{688.7 M} & 0.6827* & 0.1568 & \textbf{0.34M} & \textbf{688.5 M} & \textbf{0.8098*} & \textbf{0.4417} & \textbf{0.48M} & \textbf{977.6 M} \\
\textbf{SSR-D} & \textbf{0.7835*} & \textbf{0.3781} & 0.97M & 2.00 G & \textbf{0.6844*} & \textbf{0.1562} & 0.89M & 1.83 G & 0.8096* & 0.4425 & 1.23M & 2.53 G \\
\bottomrule
\multicolumn{13}{l}{\footnotesize * For AutoFIS, metrics refer to the re-training phase (post-pruning).} \\
\multicolumn{13}{l}{\footnotesize * The parameter count includes only the backbone network excluding embedding tables.} \\
\end{tabular}%
}
\end{table*}

To verify the robustness of SSR under different data distributions and domains, we conducted experiments on three widely used public benchmarks: Avazu, Criteo, and Alibaba. These datasets differ in feature sparsity and semantic complexity. As summarized in Table \ref{tab:public_benchmarks}, the proposed SSR framework achieves consistent improvements over all baselines across these datasets. Specifically, the dynamic variant SSR-D achieved the top performance in both AUC and LogLoss. When compared to the strongest baseline, RankMixer, SSR-D improves AUC by 0.63\% on Avazu, 0.03\% on Criteo, and 0.43\% on Alibaba. This suggests that the gains come from the model design rather than dataset-specific tuning, and transfer across benchmarks.

In addition to predictive accuracy, the static SSR-S variant demonstrates superior efficiency consistently across all benchmarks, including Avazu, Alibaba, and Criteo. Taking Avazu as a representative example, SSR-S outperforms RankMixer with an AUC of 0.7827 compared to 0.7772, yet it requires only 0.33M parameters and 688.7M FLOPs. This cuts parameters and FLOPs by roughly half relative to RankMixer, while improving AUC, showing that SSR removes redundant computation without sacrificing accuracy.

On Criteo, a highly competitive and saturated benchmark, the margins for improvement are inherently narrow. Nonetheless, SSR-S and SSR-D achieve a superior AUC of 0.8098, outperforming strong baselines like RankMixer (0.8093) and Wukong (0.8073). These results demonstrate that even in performance-saturated settings, SSR successfully identifies refined, high-order dependencies that traditional models overlook, affirming its effectiveness across diverse data environments.

\subsection{Scalability Analysis (RQ2)}\label{rq:2}

\subsubsection{Internal Efficiency Analysis}
\begin{figure}[t] 
    \centering
    \begin{subfigure}[b]{0.495\linewidth} 
        \centering
        \includegraphics[width=\linewidth]{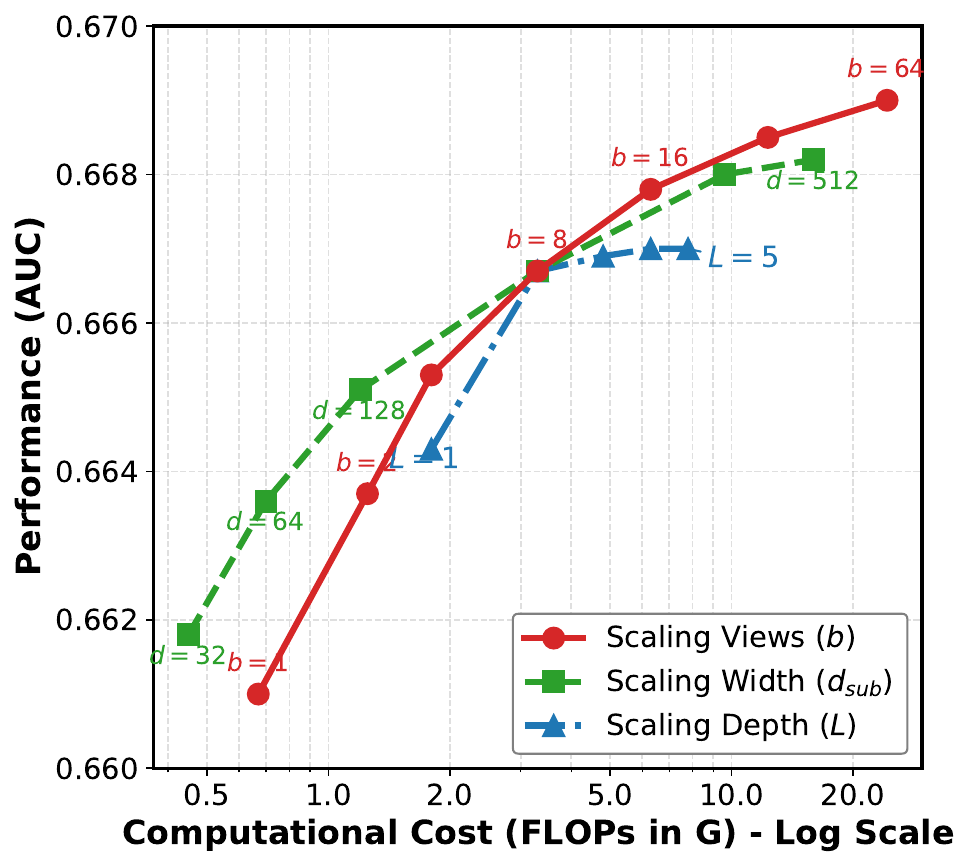}
        \caption{Industrial } 
        \label{fig:industrial}
    \end{subfigure}
    \hfill
    \begin{subfigure}[b]{0.49\linewidth}
        \centering
        \includegraphics[width=\linewidth]{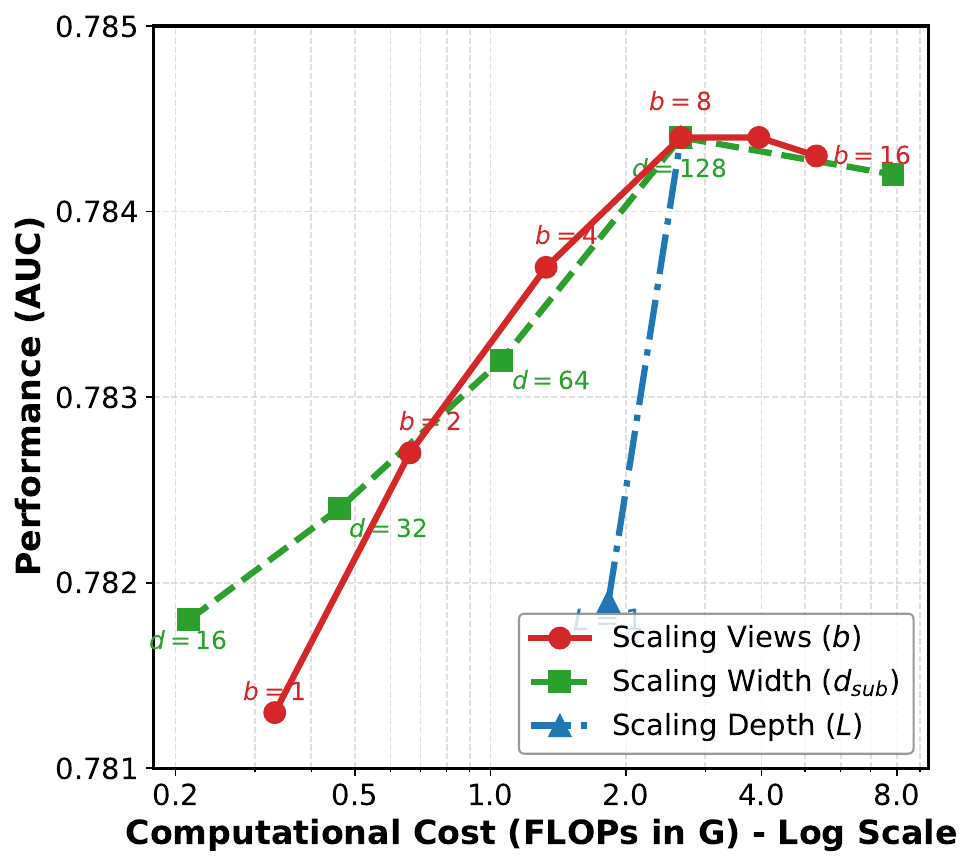}
        \caption{Avazu}
        \label{fig:avazu}
    \end{subfigure}

    \caption{Impact of scaling model dimensions on performance and cost across Industrial and Avazu datasets.}
    \Description{Impact of scaling model dimensions on performance and cost across Industrial and Avazu datasets, showing AUC and LogLoss metrics versus FLOPs.}
    \label{fig:scaling_comparison}
\end{figure}

We analyze the scaling properties in Figure \ref{fig:scaling_comparison} to identify the optimal resource allocation strategy. The results highlight that increasing the number of views ($b$) is the most reliable scaling dimension, though distinct behaviors emerge across datasets. On the smaller Avazu dataset (Figure \ref{fig:avazu}), saturation is pervasive across all dimensions. Performance gains diminish significantly as views increase from 8 to 16, and subspace width ($d_v$) even shows performance degradation beyond $d=128$. This indicates that on limited data, the model easily hits a capacity ceiling regardless of the scaling dimension.

In contrast, the billion-scale Industrial dataset (Figure \ref{fig:industrial}) exhibits a different pattern where the primary bottleneck is underfitting rather than redundancy. The performance curve for scaling views maintains a steady upward trajectory up to $b=64$ without the saturation seen in Avazu. Scaling width ($d_v$) also proves effective, serving as a strong baseline that scales well in low-to-medium resource regimes. However, it eventually exhibits diminishing returns at high complexity levels, where its curve flattens compared to the sustained growth of view scaling.

Conversely, scaling depth ($L$) consistently yields the lowest returns per FLOP on both datasets, saturating early with marginal gains. Consequently, while scaling width remains a viable secondary option, we prioritize scaling the number of views as the primary mechanism for the SSR backbone, as it offers the potential for long-term expandability on large-scale data.

\subsubsection{Scalability Efficiency Analysis}

\begin{figure}[t]
    \centering
    \includegraphics[width=0.80\linewidth]{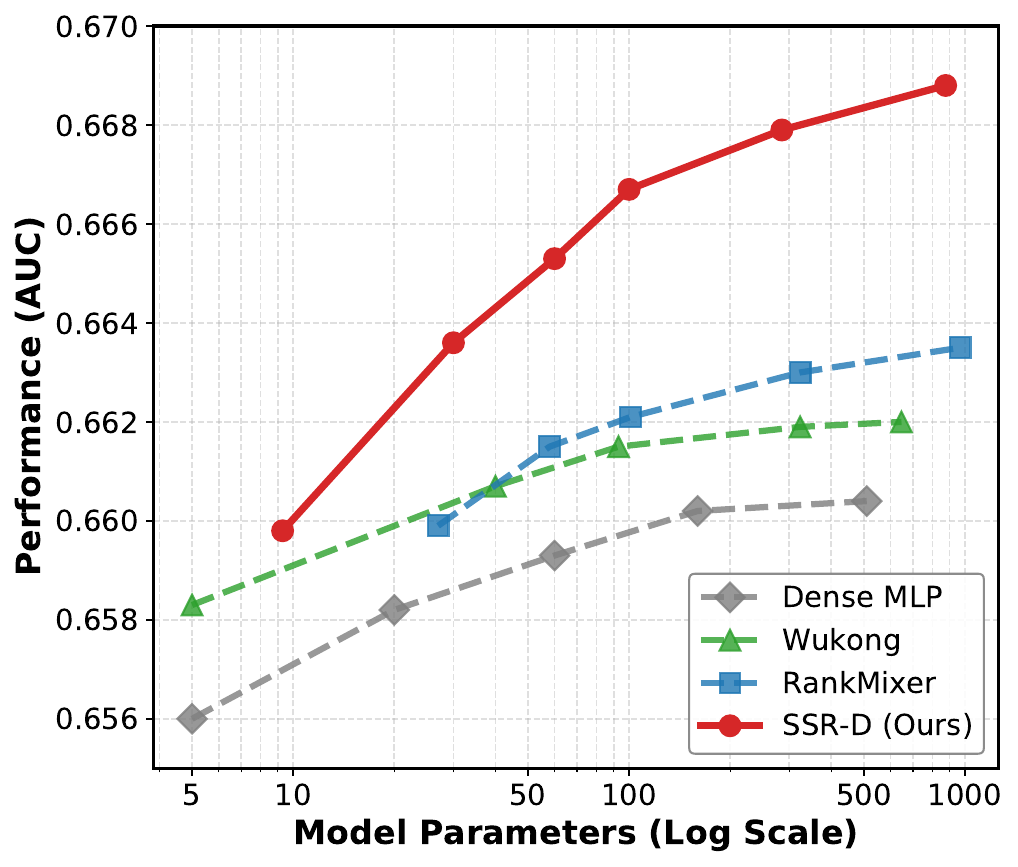}
    
    \caption{Performance (AUC) vs. Model Parameters (log scale) on the Industrial Dataset.}
    \Description{Performance (AUC) versus Model Parameters (log scale) on the Industrial Dataset comparing SSR against baselines.}
    \label{fig:parameter_comparison}
\end{figure}
We evaluate the scalability of SSR framework against two types of baselines. To ensure a rigorous comparison, we conducted an independent hyperparameter grid search for all baselines at each parameter scale. First, we compare against state-of-the-art architectures like RankMixer and Wukong to establish a strong reference point. Second, we include a standard Dense MLP to validate the structural advantage of our sparse filtering. Figure \ref{fig:parameter_comparison} plots the performance trajectory of each model across parameter scales ranging from 5M to nearly 900M.

Compared with the strongest baselines, RankMixer and Wukong, SSR exhibits not only higher accuracy but also a steeper scaling trajectory. As shown in Figure \ref{fig:parameter_comparison}, while RankMixer maintains steady improvement as parameters increase, its growth rate is flatter than that of SSR. Consequently, the performance gap between SSR and the state-of-the-art widens as the model scales up. In the large-scale increases approaching 900M parameters, SSR converts additional capacity into performance gains much more efficiently than the baselines, resulting in a larger margin. This indicates that the multi-view architecture makes better use of large-scale parameter budgets than existing methods.

Comparing our model against the Dense MLP is crucial for validating our design choices. We observed that even with carefully tuned regularization (e.g., Dropout, weight decay), the Dense MLP exhibits premature saturation, where doubling the parameter count yields diminishing returns. This plateauing effect indicates that without an explicit selection mechanism, a dense backbone struggles to utilize additional capacity to capture finer interaction patterns. In contrast, SSR maintains a steady upward trend throughout the entire scale. This confirms that the sparse filtering mechanism is pivotal for scaling. By replacing indiscriminate dense connections with selective views,  SSR allocates expanded capacity to modeling the most informative signals, thereby mitigating the saturation bottleneck that limits traditional dense networks.

\subsection{Ablation Studies \& Mechanism Analysis (RQ3)}\label{rq:3}

\subsubsection{Ablation Studies}
To validate the SSR framework, we performed comprehensive ablation studies on the Avazu and Industrial datasets. We measured the contribution of each design element by tracking the AUC performance drop relative to the SSR-D baseline, as summarized in Table \ref{tab:ablation_grouped}.

Dimension-level sparse filtering proved essential for our architecture. Eliminating this module (thereby exposing the input directly to dense blocks)leads to the most significant performance degradation, causing the AUC to decline by 0.50pt on Avazu and 0.37pt on the Industrial dataset. This sharp decline confirms our central hypothesis that globally dense connectivity is suboptimal for recommendation inputs, as forcing the backbone to process all input dimensions indiscriminately dilutes effective patterns with irrelevant connections. Complementing this, the multi-view decomposition strategy plays a vital role in maintaining model capacity. Constraining the model to a single representation subspace ($b=1$) resulted in performance losses of 0.22pt on Avazu and 0.15pt on the Industrial dataset, indicating that parallel view projections are essential for capturing diverse and complementary feature interactions.

Beyond component existence, we examined the underlying implementation mechanisms. The necessity of dynamic adaptation is evidenced by the performance drop of 0.12pt and 0.23pt when replacing the dynamic SSR-D with a static SSR-S variant, suggesting that fixed sparsity patterns fail to account for sample-specific variability. Furthermore, the superiority of our differentiable ICS operator is highlighted by its comparison with the standard Top-k selection strategy (STE), $k=d_v$. The non-differentiable nature of Top-k truncation results in a performance penalty of roughly 0.18pt, 0.29pt in AUC. In contrast, our ics provides stable gradient propagation and retains critical feature information more effectively. Finally, we replace our sparse filtering with Dropout to verify that our gains are not merely due to regularization. The resulting drastic performance drops of 0.32pt and 0.45pt demonstrate that SSR has learned meaningful sparsity.

\begin{table}[t]
\centering
\caption{\textbf{Impact of different components and mechanisms.} The baseline is the full SSR-D model. Performance changes are reported in AUC ($\times 10^{-2}$).}
\label{tab:ablation_grouped}
\begin{tabular}{l c c}
\toprule
& \multicolumn{2}{c}{\textbf{$\Delta$AUC}} \\
\cmidrule(lr){2-3} 
\textbf{Setting} & \textbf{Avazu} & \textbf{Industrial} \\
\midrule
\multicolumn{3}{l}{\textit{Component Effectiveness}} \\
\hspace{3mm} w/o Sparse Filtering & $-0.50$ & $-0.37$ \\
\hspace{3mm} w/o Multi-view Strategy & $-0.22$ & $-0.15$ \\
\midrule
\multicolumn{3}{l}{\textit{Mechanism Analysis}} \\
\hspace{3mm} Static (SSR-S) vs. Dynamic & $-0.12$ & $-0.23$ \\
\hspace{3mm} Top-k (STE) vs. ICS  & $-0.18$ & $-0.29$ \\
\hspace{3mm} Dropout vs. SSR-S & $-0.32$ & $-0.45$ \\
\bottomrule
\\
\end{tabular}
\end{table}

\subsubsection{ICS Analysis}
    


\begin{figure}[t]
  \centering
  \begin{subfigure}[t]{0.48\columnwidth}
    \centering
    \includegraphics[width=\linewidth]{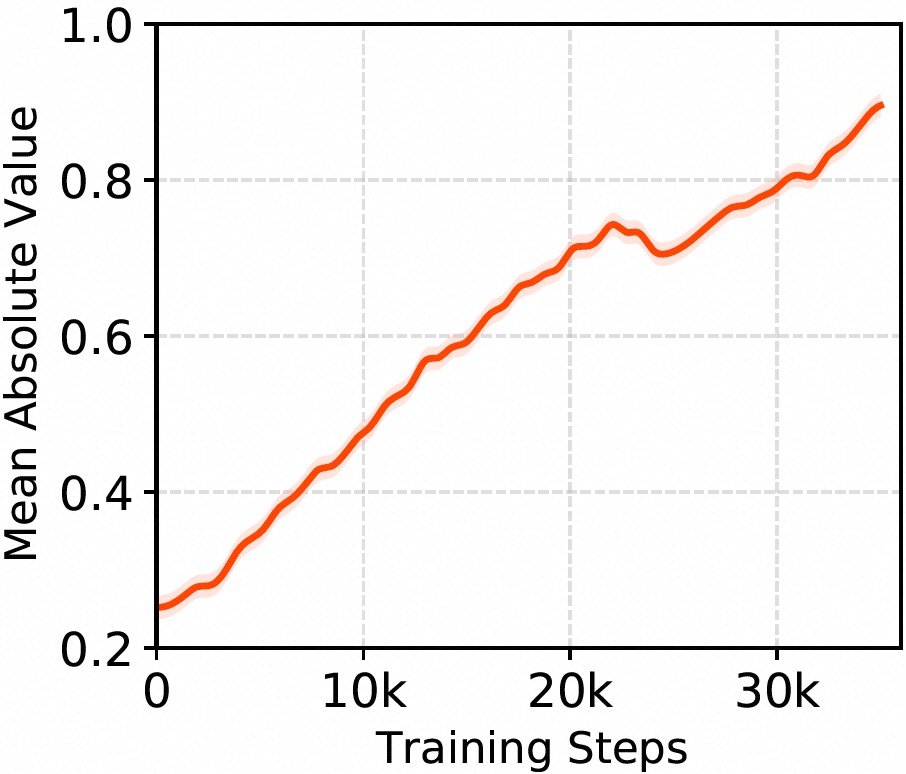}
    \caption{Layer-1 View 1 Mean Abs}
    \label{fig:a}
  \end{subfigure}\hfill
  \begin{subfigure}[t]{0.50\columnwidth}
    \centering
    \includegraphics[width=\linewidth]{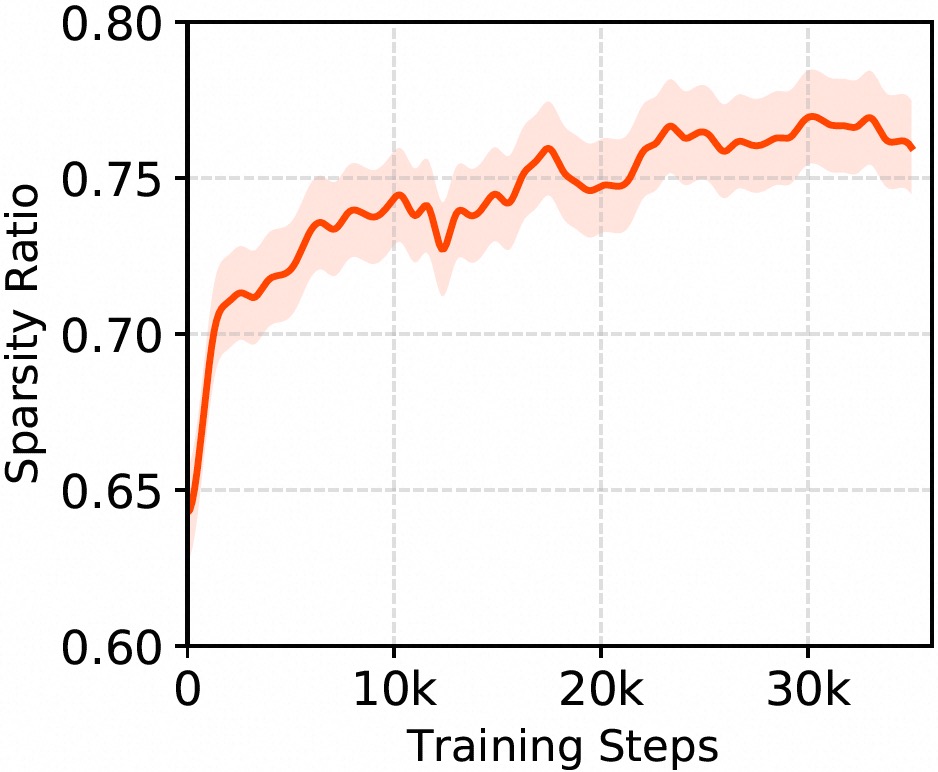}
    \caption{Layer-1 View 1 Sparsity}
    \label{fig:b}
  \end{subfigure}

  \begin{subfigure}[t]{0.49\columnwidth}
    \centering
    \includegraphics[width=\linewidth]{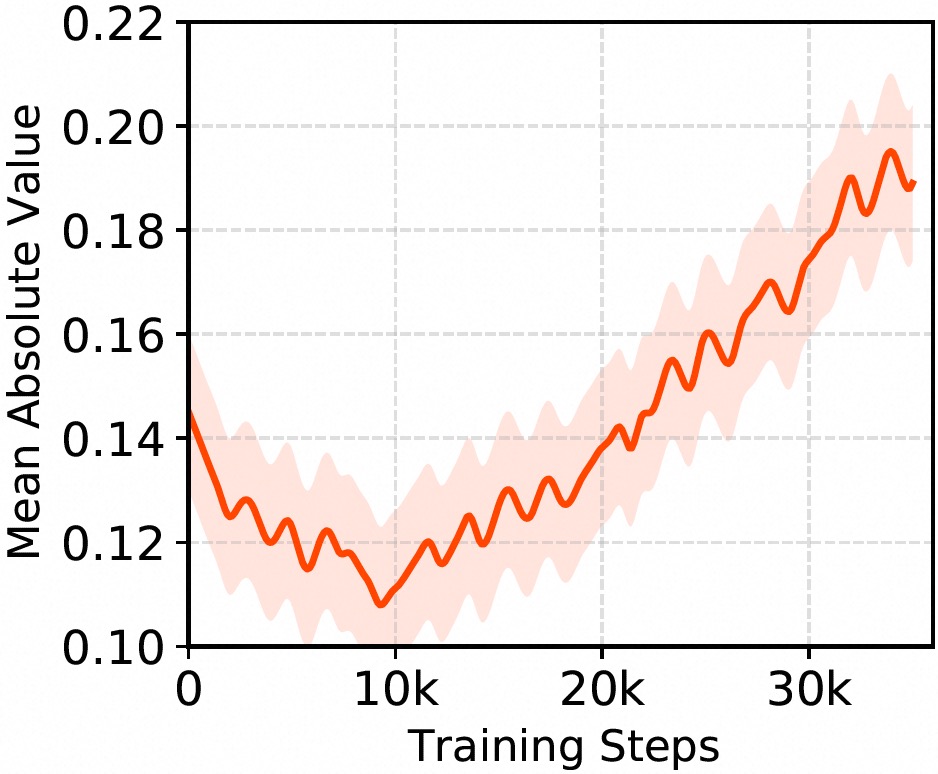}
    \caption{Layer-2 View 1 Mean Abs}
    \label{fig:c}
  \end{subfigure}\hfill
  \begin{subfigure}[t]{0.49\columnwidth}
    \centering
    \includegraphics[width=\linewidth]{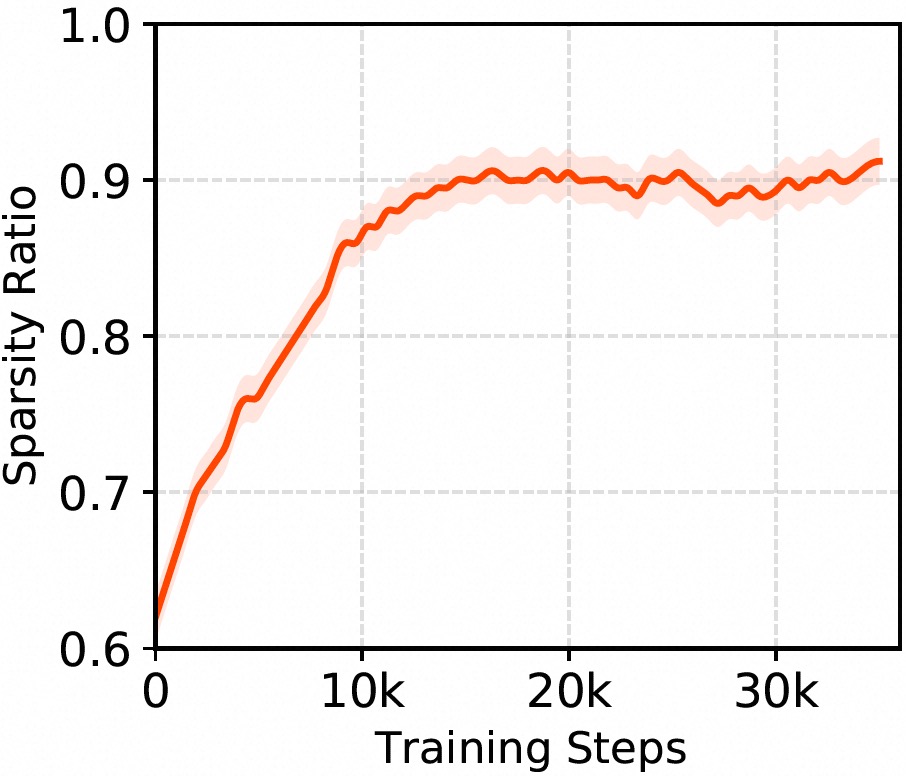}
    \caption{Layer-2 View 1 Sparsity}
    \label{fig:d}
  \end{subfigure}

  \caption{Visualization of training dynamics for the Iterative Competitive Sparse (ICS) module. }
  \Description{Visualization of training dynamics for the Iterative Competitive Sparse module showing sparsity ratios and mean absolute magnitudes across training steps for Layer 1 and Layer 2.}
  \label{fig:ici}
\end{figure}

To understand how the Iterative Competitive Sparse module learns during optimization, we visualize the first two layers in Figure \ref{fig:ici}. We tracked the sparsity ratio and the mean absolute magnitude over 35,000 steps. As shown in Figure \ref{fig:b} and \ref{fig:d}, sparsity rose quickly early on and then levels off. Layer 2 converges to a much higher sparsity (about 90\%) than Layer 1 (about 75\%), suggesting that deeper layers become more selective and produce more abstract, sparse representations. The stability observed in the later stages confirms that stable convergence rather than continual switching among feature subsets.Meanwhile, Figure \ref{fig:a} and \ref{fig:c} show that the mean absolute feature magnitude increases over training. In Layer 2, it briefly drops in the first 10,000 steps before increasing, consistent with an early suppression of weak or redundant features followed by strengthening of the remaining ones.

\begin{table}[t]
    \centering
    \caption{Sensitivity Analysis and Mechanism Validation on Avazu Dataset.}
    \label{tab:sensitivity_analysis}
    \renewcommand{\arraystretch}{1.1} 
    \setlength{\tabcolsep}{3.5mm}   
    \resizebox{\linewidth}{!}{%
    \begin{tabular}{l c c c}
        \toprule
        \textbf{Setting} & \textbf{Parameter Value} & \textbf{Sparsity (\%)} & \textbf{AUC} \\
        \midrule
        \multicolumn{4}{l}{\textit{\textbf{(A) Impact of Iterations} $T$ (with initial $\alpha_t =0.1$, w/ $\gamma$)}} \\
        \midrule
        \multirow{3}{*}{Iterations ($T$)} 
          & $T=1$ (Single Step) & 76.4\% & 0.7821 \\
          & $T=2$ & 88.6\% & 0.7826 \\
          & $\mathbf{T=5}$ \textbf{(Default)} & \textbf{91.0\%} & \textbf{0.7835} \\
        
        \midrule
        \multicolumn{4}{l}{\textit{\textbf{(B) Impact of Learnable Extinction Rates} $\alpha_t$ (with fixed $T=5$, w/ $\gamma$)}} \\
        \midrule
        \multirow{4}{*}{Extinction ($\alpha_t$)} 
          & $\alpha_t = 0.01$ & 80.4\% & 0.7832 \\
          & $\mathbf{\alpha_t = 0.1}$ \textbf{(Default)} & \textbf{91.0\%} & \textbf{0.7835} \\
          & $\alpha_t = 0.3$ & 93.3\% & 0.7833 \\
          & $\alpha_t = 0.5$ & 94.0\% & 0.7828 \\
          
        \midrule
        \multicolumn{4}{l}{\textit{\textbf{(C) Necessity of Rescaling} $\gamma$ (with fixed $T=5$, initial $\alpha_t=0.1$)}} \\
        \midrule
        \multirow{2}{*}{Rescaling ($\gamma$)} 
          & w/o $\gamma$ & 94.5\% & 0.7832 \\
          & \textbf{w/ } $\boldsymbol{\gamma}$ \textbf{(Default)} & \textbf{91.0\%} & \textbf{0.7835} \\
        \bottomrule
    \end{tabular}
    }
\end{table}

To assess the sensitivity of the ICS mechanism, we conduct a controlled grid search on Avazu (Table \ref{tab:sensitivity_analysis}), varying the number of iterations $T$, the initial extinction rate $\alpha_0$, and the rescaling factor $\gamma$. The results support the need for progressive filtering, single-step thresholding ($T=1$) yields limited sparsity and suboptimal accuracy, whereas increasing $T$ to 5 produces cleaner representations and achieves the best AUC of 0.7835 at 91.0\% sparsity. We also find that $\alpha_0$ serves as an effective sparsity regulator, smoothly shifting sparsity from 80.4\% to 94.5\% while keeping performance stable over a wide range of initial values ($\alpha_t \in [0.1, 0.5]$), indicating that the mechanism is robust rather than brittle. Finally, $\gamma$ is important for numerical stability: removing it reduces AUC to 0.7832, consistent with our analysis that explicit magnitude rescaling is needed to offset the signal attenuation.

\subsubsection{View Diversity}
\begin{figure}[t] 
    \centering
    \begin{subfigure}[b]{0.495\linewidth} 
        \centering
        \includegraphics[width=\linewidth]{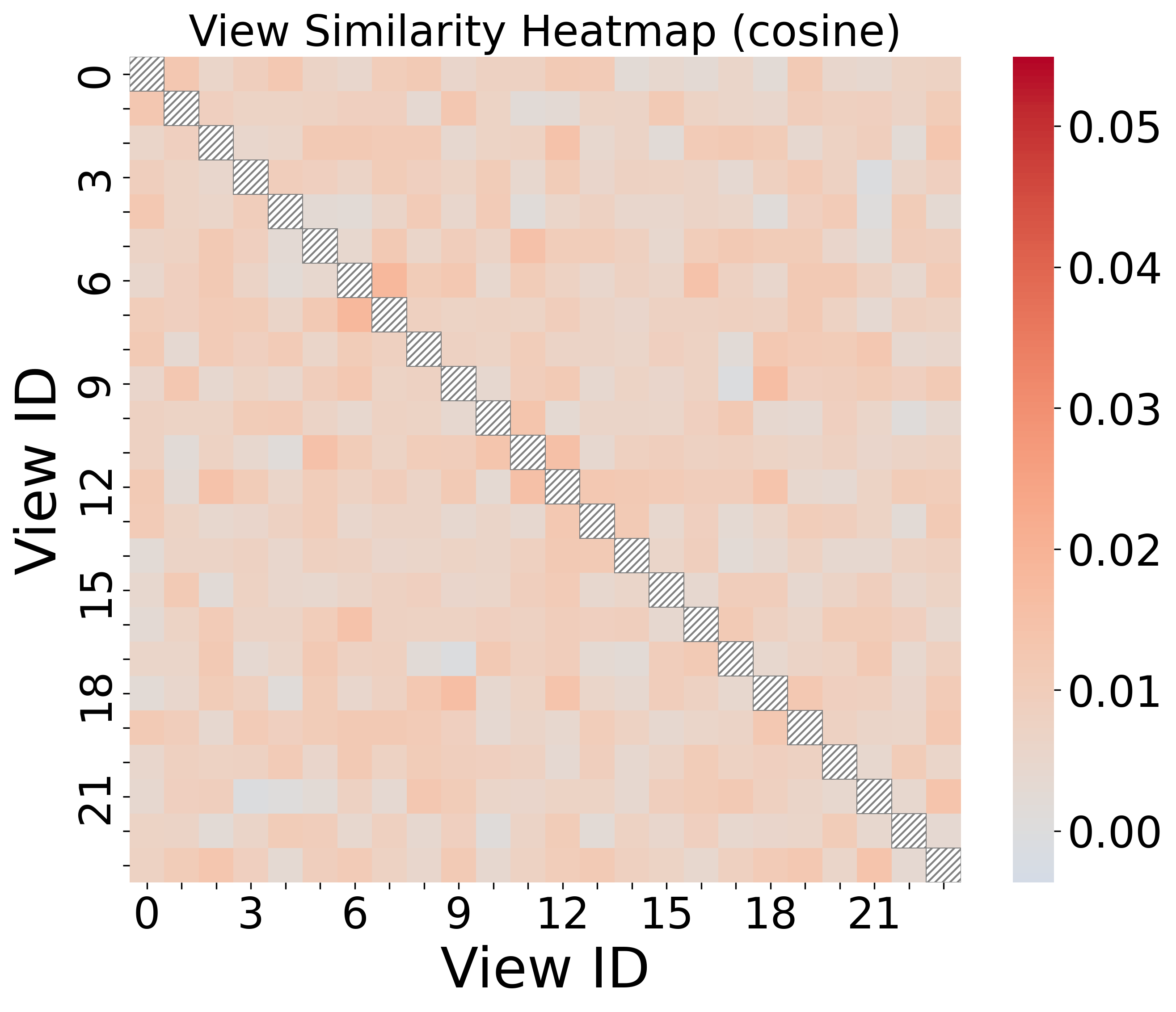}
        \caption{Layer 1 } 
        \label{fig:Layer1}
    \end{subfigure}
    \hfill
    \begin{subfigure}[b]{0.495\linewidth}
        \centering
        \includegraphics[width=\linewidth]{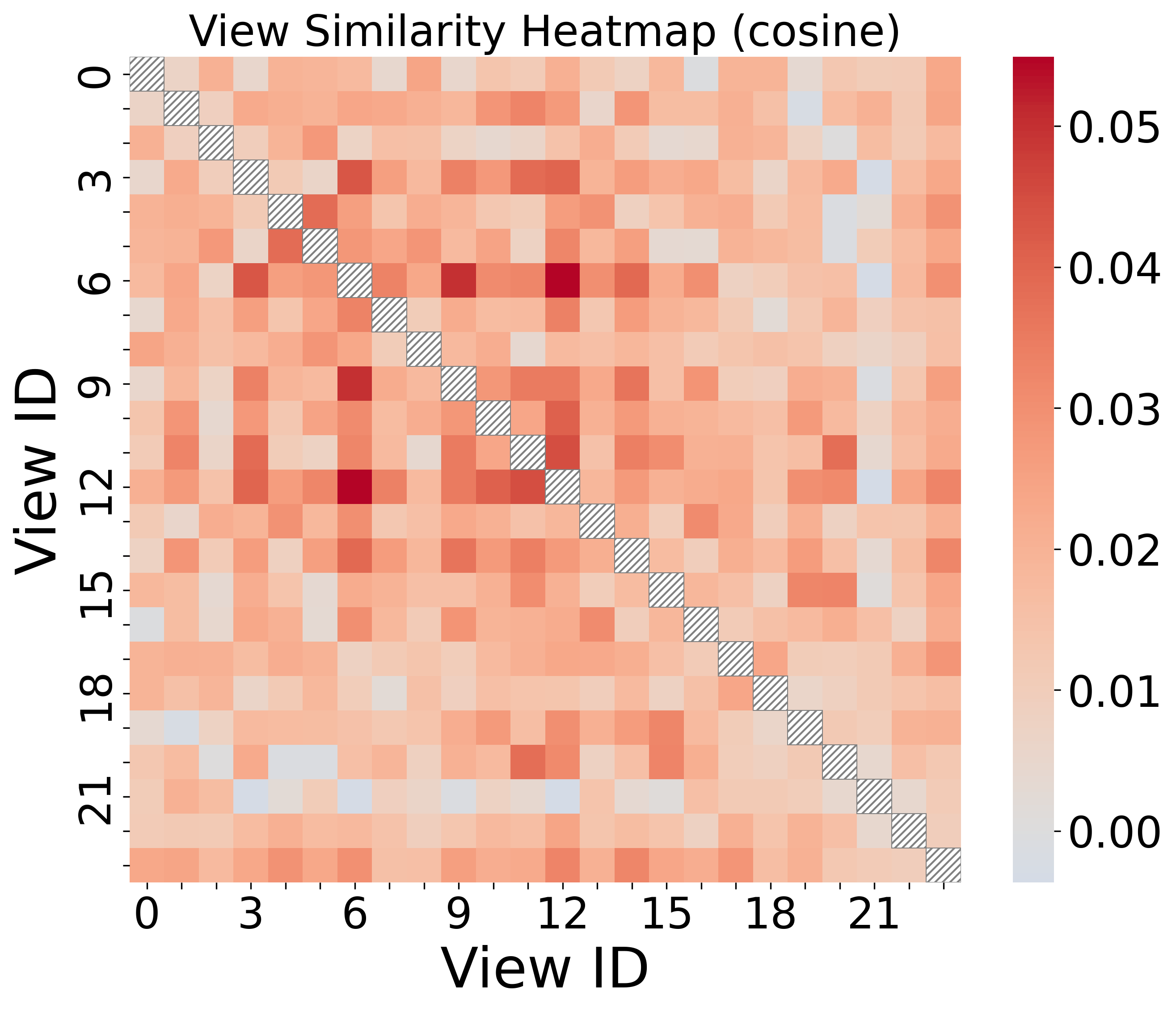}
        \caption{Layer 2}
        \label{fig:Layer2}
    \end{subfigure}

    \caption{Visualization of cosine similarity between views.}
    \Description{Visualization of cosine similarity between different views in Layer 1 and Layer 2 of the SSR framework.}
    \label{fig:similarity}
\end{figure}

To verify whether the multi-view architecture truly learns complementary patterns rather than redundant information, we visualize the pairwise cosine similarity between the projection matrix $\mathbf{W}^{proj}_{i}$ of different views in Figure \ref{fig:similarity}. The heatmaps for both Layer 1 and Layer 2 exhibit consistently low similarity scores across the off-diagonal elements. This indicates that the feature vectors generated by different views remain largely orthogonal to each other. Such a distinct separation confirms that the parallel views have successfully converged to diverse subspaces, with each view capturing a unique aspect of the feature interactions. By avoiding mode collapse where views become identical, the framework maximizes its representational capacity and ensures that the final fusion step integrates comprehensive and non-redundant signals from the input data.
SSR does not require an explicit diversity regularizer. Since all view outputs are concatenated and optimized under the same loss, training naturally suppresses redundant views and favors those that capture complementary patterns.

\subsection{Online A/B Testing (RQ4)}

\begin{table}[t]
\centering
\caption{Online A/B testing results.}
\label{tab:online_ab}
\setlength{\tabcolsep}{12pt}
\resizebox{\linewidth}{!}{%
\begin{tabular}{l c c c c}
\toprule
\multirow{2}{*}{\textbf{Model}} & \textbf{Efficiency} & \multicolumn{3}{c}{\textbf{Business Metrics (Lift)}} \\
\cmidrule(lr){2-2} \cmidrule(lr){3-5}
& \textbf{Latency} & \textbf{CTR} & \textbf{Orders} & \textbf{GMV} \\
\midrule
\textbf{SSR-D (Ours)} & 26ms(+1ms) & +2.1\% & +3.2\% & +3.5\% \\
\bottomrule
\end{tabular}%
}
\end{table}

We conducted an online A/B test in a core recommendation scenario to verify the practical value of SSR. The baseline model is RankMixer with identical parameters, which represents the current production standard. We compared this against the SSR-D over a two-week period to evaluate performance under real-world traffic. As shown in Table \ref{tab:online_ab}, SSR-D delivers consistent improvements across all key business metrics. The model achieved a 2.1\% increase in Click-Through Rate while driving substantial gains in conversion, with per capita orders rising by 3.2\% and Gross Merchandise Value by 3.5\%. These results confirm that the high-quality representations learned by SSR directly translate into better ranking decisions and higher commercial value. Crucially, these performance gains are achieved without compromising system latency. As detailed in the efficiency statistics, both the baseline RankMixer and the proposed SSR-D operate with an average response time of 25ms. This parity confirms that SSR improves recommendation quality through superior structural design rather than by increasing the inference time burden on the serving system.

\section{Related Work}
We review three lines of related work that contextualize the SSR framework: feature interaction modeling, sparsity-driven architectures, and dynamic selection mechanisms.

\subsection{From Global Dense to Sparse Filtering}
Capturing non-linear dependencies among high-dimensional sparse features is fundamental to recommender systems. Early models, like Factorization Machines, explicitly handled second-order interactions. In the deep learning era, architectures generally fall into three categories: Hybrid models (e.g., Wide\&Deep \cite{cheng2016wide}, DeepFM \cite{guo2017deepfm}) combine linear and nonlinear components to balance memorization and generalization; Self-attention mechanisms (e.g., AutoInt, AFN \cite{song2019autoint,cheng2020adaptive}) utilize multi-head attention for high-order correlations; and implicit models (e.g., DCN v2 \cite{wang2021dcn}, RankMixer \cite{zhu2025rankmixer}) rely on deep stacks of fully connected layers to capture interactions. However, a fundamental mismatch exists between these globally dense architectures and intrinsic data sparsity.  While Graph Neural Networks like IntentGC \cite{zhao2019intentgc} attempt to address sparsity by leveraging graph topology to guide interactions, they often incur costs related to graph construction and neighbor sampling in industrial settings.

Similarly, self-attention models (e.g., AutoInt \cite{song2019autoint}) theoretically capture fine-grained correlations. However, standard Softmax operations produce strictly positive weights, inherently preserving a fully connected graph. Although Sparse Attention mechanisms \cite{child2019generating} have been proposed to limit receptive fields, they often introduce complex indexing overheads. In contrast, SSR adopts a filter-then-fuse paradigm. Instead of relying on heavy graph structures or complex sparse attention indices, SSR employs explicit signal filtering. By decomposing inputs into parallel views and blocking noise before fusion, SSR enables the model to scale effectively without the saturation observed in dense baselines.
 
\subsection{From Pruning to Structural Sparsity}
To mitigate the computational burden of high-dimensional features, explicit sparsity has become an active research direction. Traditional methods largely fall into two categories: Feature Selection (e.g., AutoFIS \cite{liu2020autofis}) which prunes redundant fields, and Mixture-of-Experts (MoE) (e.g., MMOE \cite{ma2018modeling}, PLE \cite{tang2020progressive}) which uses conditional routing to expand capacity. These approaches have limitations. Feature selection often follows a model-then-prune logic—attempting to remove redundancy after dense interactions have already occurred. MoE models, while increasing capacity, face challenges with routing collapse and load balancing.
Recent advancements have shifted towards intrinsic sparsity. For instance, recent studies \cite{wang2024dynamic} propose a Dynamic Sparse Learning paradigm to train sparse models from scratch, effectively avoiding the redundancy of post-hoc pruning. Similarly, subsequent research \cite{spivsak2023scalable} utilizes sparse approximate inverses to enhance scalability in collaborative filtering autoencoders. SSR diverges from traditional post-hoc pruning and soft attention by introducing a hard-filtering paradigm. Rather than learning then deleting or preserving noise through strictly positive weights, SSR implements a learn-while-filtering mechanism from the start. Most significantly, by enforcing truncation (zero-weight connections), SSR achieves signal isolation that blocks noise propagation.

\subsection{From Gating to Global Inhibition}
To achieve input-aware adaptivity, dynamic mechanisms are essential. Existing works have explored various techniques to handle data sparsity dynamically. MaskNet \cite{wang2021masknet} and LHUC \cite{swietojanski2014learning} introduce Instance-Aware Masks to highlight informative features via element-wise gating. Other approaches leverage Locality-Sensitive Hashing \cite{chen2019improved} for efficient retrieval in edge environments or employ embedding compression \cite{kasalicky2025future} to generate sparse activations for scalable retrieval. 

However, most existing methods rely on independent gating or static projections, where feature selection decisions are made locally or via simple dot products. 
SSR advances this by proposing the Iterative Competitive Sparse (ICS) mechanism. ICS models feature selection as a dynamic system inspired by biological global inhibition. It introduces competition where dominant features suppress weaker neighbors, rather than independent gating. This allows SSR to learn a robust, global selection policy that adapts iteratively to the input context.

\section{Conclusion}
In this work, we revisited the scaling laws of recommender systems and identified the mismatch that leads to performance saturation in dense backbones. Our analysis revealed that indiscriminate mixing in standard dense layers often leads to signal dilution, necessitating a shift from passive implicit suppression to explicit signal filtering. SSR implements this paradigm through the "filter-then-fuse" topology. By employing mechanisms like Iterative Competitive Sparse (ICS), SSR blocks noise propagation at the source, ensuring that expanded model capacity is concentrated exclusively on high-SNR (Signal-to-Noise Ratio) subspaces.

Our empirical results demonstrate that this sparsity successfully breaks the scaling ceiling where dense models saturate. More broadly, this work instantiates a general principle: effective architectures align their inductive biases with the structure of their data. Just as convolutional kernels succeed because they match the spatial locality of images, and self-attention succeeds because it captures long-range dependencies in sequences, SSR's explicit sparsity succeeds because it matches the high-dimensional, subset-based interaction patterns inherent in recommendation data. This perspective challenges the prevailing reliance on globally dense connectivity and points towards a principled direction for future research: designing architectures whose structural priors reflect the sparse, combinatorial nature of user behaviors. We anticipate that explicit filtering mechanisms will be instrumental in developing larger, foundational models for recommendation that are both scalable and computationally efficient.

\begin{acks}
An AI language model was used to improve the clarity and grammar of parts of this manuscript. It was not used to generate content.
\end{acks}




\bibliographystyle{ACM-Reference-Format}
\balance                              
\bibliography{sample-base}









\end{document}